\documentclass[acmsmall]{acmart}

\usepackage{subfigure}
\usepackage{multirow}
\usepackage{algorithm}
\usepackage{algorithmic}

\acmJournal{TKDD}
\acmVolume{0}
\acmNumber{0}
\acmArticle{0}
\acmMonth{0}

\begin{document}

\title[MUSE]{MUSE: Multi-faceted Attention for Signed Network Embedding}

\author{Dengcheng Yan}
\email{yanzhou@ahu.edu.cn}
\orcid{0000-0003-1417-5269}
\author{Youwen Zhang}
\email{wen070864@gmail.com}
\orcid{0000−0003−3370−0217}
\author{Wei Li}
\email{91019@ahu.edu.cn}
\orcid{}
\author{Yiwen Zhang}
\authornote{Corresponding author.}
\email{zhangyiwen@ahu.edu.cn}
\orcid{0000-0001-8709-1088}
\affiliation{
    \department{School of Computer Science and Technology}
    \institution{Anhui University}
    \streetaddress{111 Jiulong Rd}
    \city{Hefei}
    \state{Anhui}
    \country{China}
 }

\renewcommand{\shortauthors}{Dengcheng Yan, et al.}

\begin{abstract}
Signed network embedding is an approach to learn low-dimensional representations of nodes in signed networks with both positive and negative links, which facilitates downstream tasks such as link prediction with general data mining frameworks. Due to the distinct properties and significant added value of negative links, existing signed network embedding methods usually design dedicated methods based on social theories such as balance theory and status theory. However, existing signed network embedding methods ignore the characteristics of multiple facets of each node and mix them up in one single representation, which limits the ability to capture the fine-grained attentions between node pairs. In this paper, we propose \textbf{MUSE}, a \textbf{MU}lti-faceted attention-based \textbf{S}igned network \textbf{E}mbedding framework to tackle this problem. Specifically, a joint intra- and inter-facet attention mechanism is introduced to aggregate fine-grained information from neighbor nodes. Moreover, balance theory is also utilized to guide information aggregation from multi-order balanced and unbalanced neighbors. Experimental results on four real-world signed network datasets demonstrate the effectiveness of our proposed framework.
\end{abstract}



\maketitle

\section{Introduction}

Signed relations such friend/foe and trust/distruct are ubiquitous in many real-world social networks, including both online and offline social networks, which can be uniformly modeled as signed networks with positive and negative links ~\cite{kunegis2009slashdot}. Earlier social psychologist have been devoted to the development of signed network analysis based on social theories such as balance theory ~\cite{heider1946attitudes}.

\begin{figure}[htbp]
    \centering
    \includegraphics[scale=1.0]{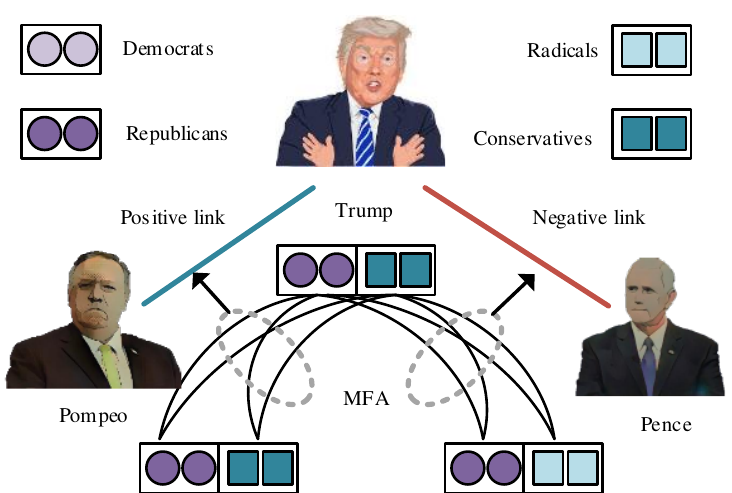}
    \caption{Motivation example.}
    \label{fig:motivation}
\end{figure}

Recently, with the wide adoption of data mining and machine learning from various domains, techniques for signed network analysis have evolved from observations, measurements to mining tasks \cite{tang2016survey,huang20motif,zhao2020network}. Among these techniques, signed network embedding ~\cite{wang2017signed,kim2018side,derr2018signed,huang2019signed,li2020learning}, which learns a low-dimension vector representation for each node in signed networks, is one of the most promising one for it can automatically extract features and can employ various kinds of data mining and machine learning models to perform signed network analysis tasks such as link sign prediction.

However, conventional unsigned network embedding methods cannot be directly applied to signed network because it has been proven that negative links in signed networks are extremely valuable than positive links since the existence of a little fraction of negative links can boost the performance for recommendation ~\cite{ma2009learning} and link sign prediction ~\cite{jung2016personalized}. Signed network embedding recently attracts an increasing attention. Some earlier approaches derive normalized spectral analysis ~\cite{zheng2015spectral}, adopt the log-bilinear model ~\cite{javari2020rose}, or treat both positive and negative neighbors the same during the aggregation process\cite{lee2020asine}. Later, graph neural networks \cite{kipf2017gcn} are introduced \cite{derr2018signed,li2020learning} to aggregate information from neighbor nodes.

Despite of the numerous methods proposed for signed network embedding, none of them, as far as we know, takes a fine-grained look at the multiple facets of nodes in signed networks and their influence on signed link formation, while actually this exists ubiquitously in various domains. Taking Figure \ref{fig:motivation} as an example, different types of links formed between the three famous politicians in the U.S. Pompeo and Trump setup a positive link because of similar party affiliation and political attitudes while Pence and Trump setup a negative link because of different political attitudes despite of similar party affiliation.

To take a fine-grained look at the multiple facets of nodes in signed networks and reveal the rules of signed link formation on the basis of multiple facets characteristics, we propose a \textbf{MU}lti-faceted attention-based \textbf{S}igned network \textbf{E}mbedding framework, named \textbf{MUSE}. It divides the embedding of a node into multiple facets and employs multi-faceted attention (MFA), a joint intra- and inter-facet attention mechanism, to aggregate fine-grained information from multi-order neighbor nodes. Our main contributions are summarized as follows:

\begin{itemize}
    \item We propose a multi-faceted attention-based signed network embedding framework to capture fine-grained attentions between node pairs. Attentions are calculated from both intra- and inter-facet views.
    \item We design a multi-order neighbor aggregation mechanism to capture high-order relations among nodes in signed network. Each order of relations is also divided into balanced and unbalanced types based on extended balance theory. 
    \item We conduct extensive experiments on four real-world signed network. The results demonstrate the effectiveness of our propose framework.
\end{itemize}

\section{Preliminaries}

\begin{definition}[Signed network ~\cite{tang2016survey}]
A signed network is denoted as $G = \left(V, E^{+}, E^{-}\right)$, where $V = \{{v_{1}, v_{2}, \cdots, v_{N}}\}$ denotes the set of nodes, and $E^{+}$ and $E^{-}$ are the sets of positive and negative links, respectively. 
\end{definition}

A signed network is usually represented by an adjacency matrix $A$ ($A \in \{-1,0,1\}^{|V| \times |V|}$), where $A_{ij}=1,-1,0$ denotes a positive, negative and no link between node $v_{i}$ and node $v_{j}$, respectively.

\begin{figure}[h]
    \centering
    \includegraphics[scale=0.6]{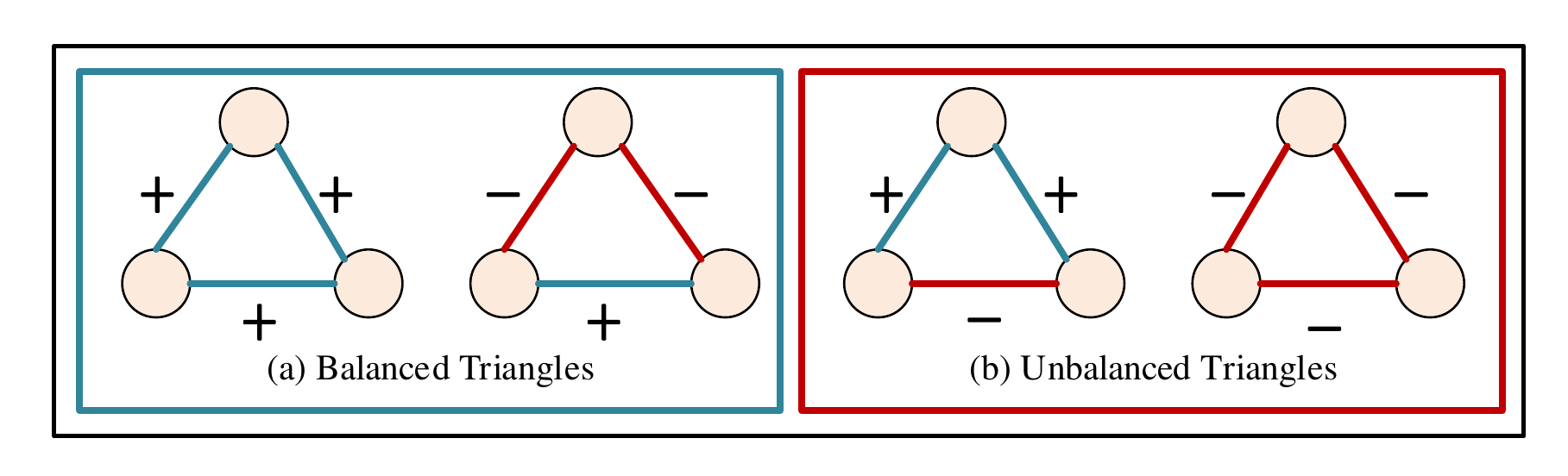}
    \caption{Four types of undirected triangles in signed network.}
    \label{fig:balance-un-triangles}
\end{figure}

Social scientists were the first to study the characteristics of signed networks under the guidance of balance theory \cite{heider1946attitudes,cartwright1956structural} which can be depicted as “a friend of my friend is my friend” and “an enemy of my friend is my enemy”. Figure \ref{fig:balance-un-triangles} shows four types of triangles in signed network, where the product of the three links of balanced triangles is positive while it is negative for unbalanced triangles. While balanced triangles in social networks are more reasonable and exist more widely, unbalanced triangles do exist exactly with a noticeable proportion. Thus information from unbalanced neighborhood need to be taken into account specially during signed network embedding. 

To be more general, we first define the higher-order balanced and unbalanced neighbors as the basis for neighbor information aggregation in signed network embedding.

\begin{definition}[Higher-order neighbors \cite{derr2018signed}]
\label{def:higher-order-neighbors}
The $l$-th order neighbors of node $v_{i}$ is defined as the set of nodes within $l$-hop from node $v_{i}$, denoted as $N_{i}(l) = \{v_{j} | A_{ij} \neq 0 \} \cup \{ v_{j} | A_{ik_{1}} A_{k_{1}k_{2}} \cdots A_{k_{l-1}j} \neq 0, l \in \mathbb{N}, l>1\} \left(l \in \mathbb{N}, l \geq 1 \right)$ . It can be divided into the $l$-th order balanced neighbors $B_{i}(l) = \{v_{j} | A_{ij} = 1 \} \cup \{ v_{j} | A_{ik_{1}} A_{k_{1}k_{2}} \cdots A_{k_{l-1}j} = 1, l \in \mathbb{N}, l>1\} \left(l \in \mathbb{N}, l \geq 1 \right)$ and the $l$-th order unbalanced neighbors $U_{i}(l) = \{v_{j} | A_{ij} = -1 \} \cup \{ v_{j} | A_{ik_{1}} A_{k_{1}k_{2}} \cdots A_{k_{l-1}j} = -1, l \in \mathbb{N}, l>1\} \left(l \in \mathbb{N}, l \geq 1 \right)$.
\end{definition}

\section{Proposed Framework - MUSE}

\begin{figure}[h]
    \centering
    \includegraphics[scale=0.75]{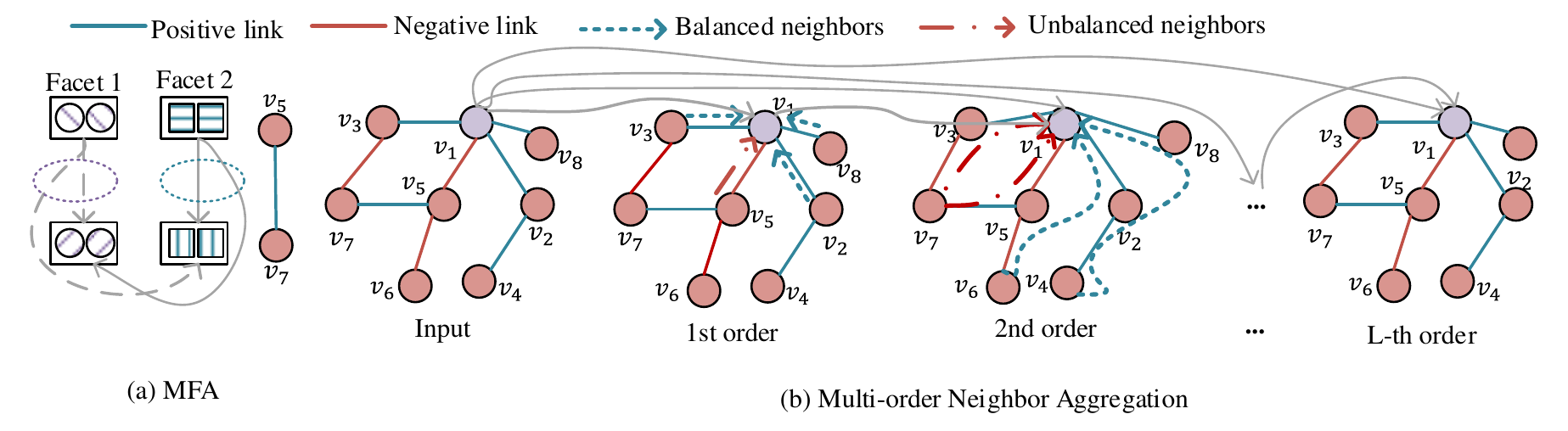}
    \caption{The framework of MUSE.}
    \label{fig:framework}
\end{figure}

The overview of the MUSE framework is shown in Figure \ref{fig:framework}. In this framework, the embedding of each node is divided into multiple facets and the specially designed multi-faceted attention (MFA) is employed to aggregate information from multi-order neighbors. It consists of two main components: 1) \textbf{multi-faceted attention}, which weights the different importance of each facet of a neighbor node's features from a fine-grained perspective. 2) \textbf{multi-order neighbor aggregation}, which aggregates information from multi-order balanced and unbalanced neighbors. It is worth noting that multi-order neighbor aggregation is different from multiple GCN layers. In the MUSE framework, all orders of neighbor aggregations are actually in the same layer and use a common input. Of course multiple layers of multi-order neighbor aggregation can be chained as GCN does.

\subsection{Multi-faceted Attention}

In the MUSE framework, the embedding $\mathbf{h}_{i}$ of node $i$ in a signed network $G$ is represented in a fine-grained way by dividing it into $M$ facets, that is $\mathbf{h}_{i} = [\mathbf{h}_{i1}, \mathbf{h}_{i2}, \cdots, \mathbf{h}_{im}, \cdots, \mathbf{h}_{iM}]$ where $\mathbf{h}_{im} \in \mathbb{R}^{D}$, $\mathbf{h}_{i} \in \mathbb{R}^{M\times D}$ and $D$ is the dimension of each facet embedding of a node. Specifically, each facet of the embedding captures an aspect of latent features.

To model the fine-gained relations between facets, a multi-faceted attention (MFA) mechanism is introduced. MFA consists of two parts: the intra-facet attention coefficient calculation and the attention-based inter-facet neighborhood information aggregation.

The intra-facet attention coefficient calculation calculates an attention coefficient that indicates the importance of the facet $m$ of node $j$ to node $i$'s representation of corresponding facet by considering intra-facet relations, as illustrated in Equation (\ref{eq:intra-facet-att-coe-cal}),

\begin{equation}
    \label{eq:intra-facet-att-coe-cal}
    e_{ijm}=\sum_{s=1}^{M} \operatorname{ATT} \left(\mathbf{W}_{T}\mathbf{h}_{im},  \mathbf{W}_{T} \mathbf{h}_{js}\right)
\end{equation}
where $e_{ijm}$ is the attention coefficient of node $i$ on node $j$ in facet $m$, $\mathbf{W}_{T} \in \mathbb{R}^{D \times D}$ is a weight matrix for linear transformation, and $\operatorname{ATT}$ is an operation to compute the attention coefficient.

To make the attention coefficient comparable across different facets, normalization is adopted across all facets. In the MUSE framework, we employ softmax to normalize the attention coefficient, as shown in Equation (\ref{eq:intra-facet-att-coe-cal-norm}),

\begin{equation}
    \label{eq:intra-facet-att-coe-cal-norm}
    \alpha_{ijm}= \operatorname{softmax} \left( e_{ijm} \right) = \frac{\exp \left( e_{ijm} \right)}{\sum_{s=1}^{M} \exp \left( e_{ijs} \right)}
\end{equation}

According to \cite{velivckovic2017gat}, in the MUSE framework, $\operatorname{ATT}$ operation is implemented using a single-layer feedforward neural network with parameters $\mathbf{W}_{A} \in \mathbb{R}^{2D}$ and nonlinear activation function $\operatorname{LeakyReLU}$, as shwon in Equation (\ref{eq:intra-facet-att-coe-cal-impl}),

\begin{equation}
    \label{eq:intra-facet-att-coe-cal-impl}
    \alpha_{ijm} = \frac{ \sum_{s=1}^{M} \exp \left( \sigma \left( \mathbf{W}^{T}_{A} [\mathbf{W}_{T} \mathbf{h}_{im} \| \mathbf{W}_{T} \mathbf{h}_{js}] \right) \right)}{\sum_{m=1}^{M} \sum_{s=1}^{M} \exp \left( \sigma \left( \mathbf{W}^{T}_{A} [\mathbf{W}_{T} \mathbf{h}_{im} \| \mathbf{W}_{T} \mathbf{h}_{js}] \right) \right)}
\end{equation}
where $\|$ is the concatenation operation and $\sigma$ is the activation function LeakyReLU.

Then the attention-based inter-facet neighborhood information aggregation use the attention coefficient calculated by Equation (\ref{eq:intra-facet-att-coe-cal-impl}) to aggregate information of each facet from a node's neighbors, as shown in Equation (\ref{eq:att-inter-facet-agg-facet}),

\begin{equation}
    \label{eq:att-inter-facet-agg-facet}
    \mathbf{h}_{im} = \sum_{v_{j} \in \mathcal{N}(v_{i})} \alpha_{ijm} \mathbf{h}_{jm}
\end{equation}
where $\mathcal{N}(v_{i})$ is the neighbor set of node $v_{i}$.

\subsection{Multi-order Neighbor Aggregation}

To capture both direct and higher-order relations between node pairs, we design a multi-order neighbor aggregation mechanism. $l$-th order neighbor aggregation aggregates information with multi-faceted attention from a node's $l$-th order neighbors as defined in Definition \ref{def:higher-order-neighbors} and combines the output of $(l-1)$-th order neighbor aggregation as well. Different from unsigned network embedding, the neighbor of a node in signed network consists of both balanced and unbalanced neighbors, both of which are important. But according to balance theory, the number of balanced neighbors are usually significantly more than unbalanced neighbors. To preserve the significance of unbalanced neighbors, the process of both attention and aggregation is divided into to two parts, as shown in Equation (\ref{eq:multi-order-agg-balanced}) and (\ref{eq:multi-order-agg-unbalanced}),

\begin{equation}
    \label{eq:multi-order-agg-balanced}
    \mathbf{h}_{i}^{B(l)} = \mathbf{h}_{i}^{BU(l-1)} + \sum_{v_{j} \in B_{i}(l)} \left[ \alpha_{ij1}^{B(l)} \mathbf{h}_{j1}^{0} \| \cdots \| \alpha_{ijM}^{B(l)} \textbf{h}_{jM}^{0}\right]
\end{equation}

\begin{equation}
    \label{eq:multi-order-agg-unbalanced}
    \mathbf{h}_{i}^{U(l)} = \mathbf{h}_{i}^{BU(l-1)} + \sum_{v_{j} \in U_{i}(l)} \left[ \alpha_{ij1}^{U(l)} \mathbf{h}_{j1}^{0} \| \cdots \| \alpha_{ijM}^{U(l+1)} \textbf{h}_{jM}^{0}\right]
\end{equation}
where $\mathbf{h}_{i}^{B(l)}$ and $\mathbf{h}_{i}^{U(l)}$ are the $l$-th order balanced and unbalanced representations of node $i$, respectively. $h_{j1}^{0}$ is the original input features of facet $1$ of node $j$. $\alpha_{ij1}^{B(l)}$ and $\alpha_{ij1}^{U(l)}$ are the attention coefficients on facet $1$ of the $l$-th order balanced and unbalanced neighbor node $j$, respectively, which can be computed according to Equation (\ref{eq:intra-facet-att-coe-cal-impl}). $\mathbf{h}_{i}^{BU(l)}$ is the composite embedding of node $i$ from the output of $l$-th order neighbor aggregation by a nonlinear transformation of the concatenation of both balanced and unbalanced representations, as shown in Equation (\ref{eq:multi-order-agg-composite}), 

\begin{equation}
    \label{eq:multi-order-agg-composite}
    \mathbf{h}_{i}^{BU(l)} = tanh(\mathbf{W_{BU}}^{T}[\mathbf{h}_{i}^{B(l)} \| \mathbf{h}_{i}^{U(l)}])
\end{equation}
where $\mathbf{W_{BU}} \in \mathbb{R}^{2MD \times MD}$ is the transformation matrix and $tanh$ is the nonlinear activation fucntion. Here we set $\mathbf{h}_{i}^{BU(0)}$ equal to $\mathbf{h}_{i}^{0}$ which is the initial embedding of node $i$.

\subsection{Optimization}

\begin{algorithm}
  \caption{Embedding Generation Process of MUSE}
  \label{algo:muse}
  \begin{algorithmic}[1]
    \REQUIRE Signed network $G = \left(V, E^{+}, E^{-}\right)$; number of facets $M$; dimension of each facet $D$; initial embedding $\{ \mathbf{h}_{i}^{0} = \left[\mathbf{h}_{i1}^{0}, \cdots, \mathbf{h}_{iM}^{0} \right], \forall v_{i} \in V \}$; number of orders $L$; weight matrices $W_{T}$, $W_{A}$ and $W_{BU}$; logistic regression parameters $W_{L}$ and $b$
    \ENSURE Node embedding $\mathbf{h}_{i}, \forall v_{i} \in V $
    \STATE Set $\mathbf{h}_{i}^{BU(0)}=\mathbf{h}_{i}^{0}$
    \FOR{$v_{i} \in V$}
        \FOR{$l$ = 1 to $L$}
            \FOR{$m$ =1 to $M$}
                \STATE Calculate balanced attention coefficient $\alpha_{ijm}^{B(l)}$ using Equation (\ref{eq:intra-facet-att-coe-cal-impl});
                \STATE Calculate unbalanced attention coefficient $\alpha_{ijm}^{U(l)}$ using Equation (\ref{eq:intra-facet-att-coe-cal-impl});
            \ENDFOR
            \STATE Calculate balanced embedding $\mathbf{h}_{i}^{B(l)}$ using Equation (\ref{eq:multi-order-agg-balanced});
            \STATE Calculate unbalanced embedding $\mathbf{h}_{i}^{U(l)}$ using Equation (\ref{eq:multi-order-agg-unbalanced});
            \STATE Calculate the composite embedding $\mathbf{h}_{i}^{BU(l)}$ using Equation (\ref{eq:multi-order-agg-composite});
        \ENDFOR
        \STATE Update $W_{T}$, $W_{A}$, $W_{BU}$, $W_{L}$ and $b$ using Adam;
    \ENDFOR
    \STATE Set $\mathbf{h}_{i} = \mathbf{h}_{i}^{BU(L)}, \forall v_{i} \in V $;
    \RETURN $\mathbf{h}_{i}, \forall v_{i} \in V $
  \end{algorithmic}
\end{algorithm}

In the MUSE framework, the embedding of each node in signed networks can be generated by aggregating information from multi-order balanced and unbalanced neighbors with multi-faceted attention. Usually the goal of network embedding is to preserve as much structure information in the embedding space as in the original network space. However, in signed network, the link sign is also important to reveal the relationships between nodes. Thus, in the MUSE framework, the objective function consists of two parts: structure-preserving loss and link sign classification loss.

Followed by the extended structural balance theory ~\cite{cygan2012sitting}, which conveyed the basic idea for a user should be able to have their  “friends” closer than their “foes". The goal of structure-preserving loss $L_{ST}$ is to have node pairs with positive links to reside closer while node pairs with negative links to reside farther in the embedding space, which can be formulated as Equation (\ref{eq:loss-struct}).

\begin{equation}
    \label{eq:loss-struct}
    L_{ST}=\frac{1}{\left|E^{+}\right|} \sum_{e_{ij} \in E^{+}} \left\|\textbf{h}_{i}-\textbf{h}_{j}\right\|_{2}^{2}-\frac{1}{\left|E^{-}\right|} \sum_{e_{ik} \in E^{-}} \left\|\textbf{h}_{i}-\textbf{h}_{k}\right\|_{2}^{2}
\end{equation}

The goal of link sign classification loss $L_{SN}$ is to minimize the error of link sign prediction. In the MUSE framework, we choose logistic regression as the link sign prediction model, which is formulated as Equation (\ref{eq:link-sign-pred-l-r}),

\begin{equation}
    \label{eq:link-sign-pred-l-r}
    \hat{y}_{ij} = \operatorname{sigmoid} \left(\mathbf{W}_{L} \left[ \mathbf{h}_{i} \| \mathbf{h}_{j} \right] + \mathbf{b} \right)
\end{equation}
where $\mathbf{W}_{L} \in \mathbb{R}^{2MD}$ and $b \in \mathbb{R}$ are parameters for logistic regression model. Then $L_{SN}$ can be formulated as Equation (\ref{eq:loss-sign-pred}).

\begin{equation}
    \label{eq:loss-sign-pred}
    L_{SN}=-\frac{1}{|E|} \sum_{e_{i j \in E}} \left(y_{i j}\cdot \log \hat{y}_{ij} +(1 - y_{i j}) \cdot \log (1 - \hat{y}_{ij}) \right)
\end{equation}
where $\sigma$ is an activation function. The ground-truth $y_{ij} =1$ if $\textbf{A}_{i j}=1$ otherwise $y_{ij} = 0$.

\begin{table}[!htp]
	\centering
	\begin{tabular}{lrrrr}
		\toprule
		 Dataset           & Bit.Alpha & Bit.OTC   & Slashdot    & Epinions \\
		\midrule
		 \# Users           & 3,775     & 5,875     & 37,626      & 45,003       \\
		 \# Positive  & 12,721    & 18,230    & 313,543     & 513,851   \\
         \# Negative  & 1,399     & 3,259     & 105,529     & 102,180  \\
		\bottomrule           
	\end{tabular}
	\caption{The statistics of datasets.}
	\label{tab:datasets-stat}
\end{table}

The overall objective function $L$ is then balanced between $L_{ST}$ and $L_{SN}$ with a weight $\lambda$ as shown in Equation (\ref{eq:loss-total}), and Adam \cite{kingma2014adam} is used as the optimizer in MUSE. The detailed procedure is described in Algorithm \ref{algo:muse}.

\begin{equation}
    \label{eq:loss-total}
    L = L_{ST} + \lambda L_{SN}
\end{equation}

\section{Experiments}

The effectiveness of our proposed framework is evaluated on four real-world datasets \cite{li2020learning}, Bitcoin-Alpha,  Bitcoin-OTC, Slashdot and Epinions, against the state-of-the-art signed network embedding methods.

\subsection{Datasets}

The datasets used in the experiment are four real-world signed networks. Bitcoin-Alpha and Bitcoin-OTC are online markets where users conduct transactions anonymously by using Bitcoins and Bitcoin accounts. In these markets, users can label other users as trust or distrust to alleviate risky transactions from scammers. The trust and distrust relations between users are then modeled as positive and negative links in a signed network, respectively. Slashdot is a technology news site where users can make friends or foes with others. The relations between friends and between foes are modeled as positive and negative links, respectively. Epinions is an online review site where users can seek or post reviews about some product, and a user can be labeled as trust or distrust by other users based on their assessments on his/her reviews. The statistics of the four datasets is shown in Table \ref{tab:datasets-stat}.

\subsection{Baselines}

We evaluate the effectiveness of our proposed framework against state-of-the-art signed network embedding methods by performing link sign prediction task with the node embeddings output by them and then comparing the F1 and AUC metrics of them. The baselines are list as follows:

\begin{itemize}
    \item SiNE ~\cite{wang2017signed}: It utilizes deep learning technique to obtain the node embeddings under the guidance of extended structural balance theory.
    \item SIDE ~\cite{kim2018side}: It distinguishes the directions of signed links and adopts a likelihood optimization over both direct and indirect signed connections to obtain the node embeddings.
    \item SGCN ~\cite{derr2018signed}: It employs graph convolutional layers to aggregate information from two separated sets of neighbors, i.e., balanced neighbors and unbalanced neighbors, to generate both balanced and unbalanced embeddings of a node simultaneously.
    \item SiGAT ~\cite{huang2019signed}: It generalizes graph attention network to signed networks and incorporates graph motifs to obtain the node embeddings under both balance theory and status theory.
    \item SNEA ~\cite{li2020learning}: It utilizes the masked self-attentional layers to aggregate more important information from neighbor nodes to generate the node embeddings.
\end{itemize}

\begin{table*}[!htp]
	\centering
	\begin{tabular}{lcrrrrrrrr}
		\toprule
		 Dataset                     & Metric& SiNE      & SIDE      & SGCN      & SiGAT     & SNEA     & MUSE-1     & MUSE           & Improv.(\%)      \\
		\midrule
		 \multirow{2}{*}{Bit.Alpha}  & F1     & 0.895     & 0.753     & 0.915     & 0.894     & 0.927    & 0.950      & \textbf{0.952} & 2.70 \\
		                            & AUC    & 0.781     & 0.642     & 0.801     & 0.775     & 0.816    & 0.808      & \textbf{0.854} & 4.66 \\

		 \multirow{2}{*}{Bit.OTC}    & F1     & 0.876     & 0.728     & 0.908     & 0.903     & 0.924    & 0.928      & \textbf{0.933} & 0.97\\
		                            & AUC    & 0.782     & 0.632     & 0.804     & 0.796     & 0.818    & 0.809      & \textbf{0.855} & 4.52\\

		 \multirow{2}{*}{Slashdot}   & F1     & 0.850     & 0.624     & 0.859     & 0.857     & 0.868    & 0.872      & \textbf{0.885} & 1.96\\
		                            & AUC    & 0.785     & 0.554     & 0.786     & 0.789     & 0.799    & 0.813      & \textbf{0.855} & 7.01\\

		 \multirow{2}{*}{Epinions}   & F1     & 0.902     & 0.725     & 0.920     & 0.917     & 0.933    & 0.932      & \textbf{0.943} & 1.07\\
		                            & AUC    & 0.831     & 0.617     & 0.849     & 0.853     & 0.861    & 0.867      & \textbf{0.910} & 5.69\\
		\bottomrule           
	\end{tabular}
	\caption{Link sign prediction results with F1 and AUC. The last column is the improvement of MUSE over the best baseline except MUSE-1}
	\label{tab:overall-performance}
\end{table*}

\begin{figure}
	\centering
	\subfigtopskip=0pt
	\subfigure[Bit.Alpha - AUC]{
		\includegraphics[scale=0.45]{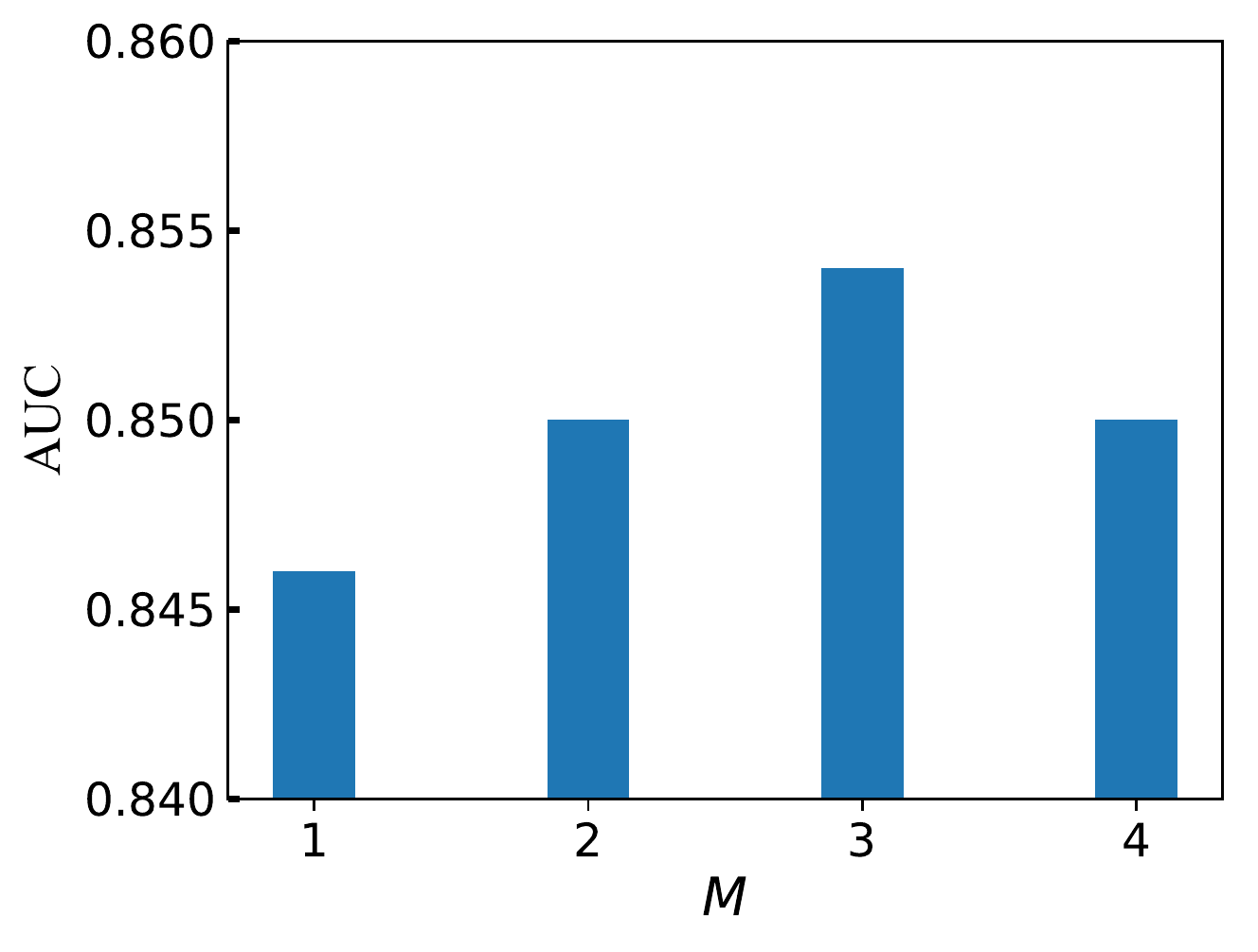}
		\label{fig:effect-facet-alpha-auc}
	}
	\subfigure[Bit.Alpha - F1]{
		\includegraphics[scale=0.45]{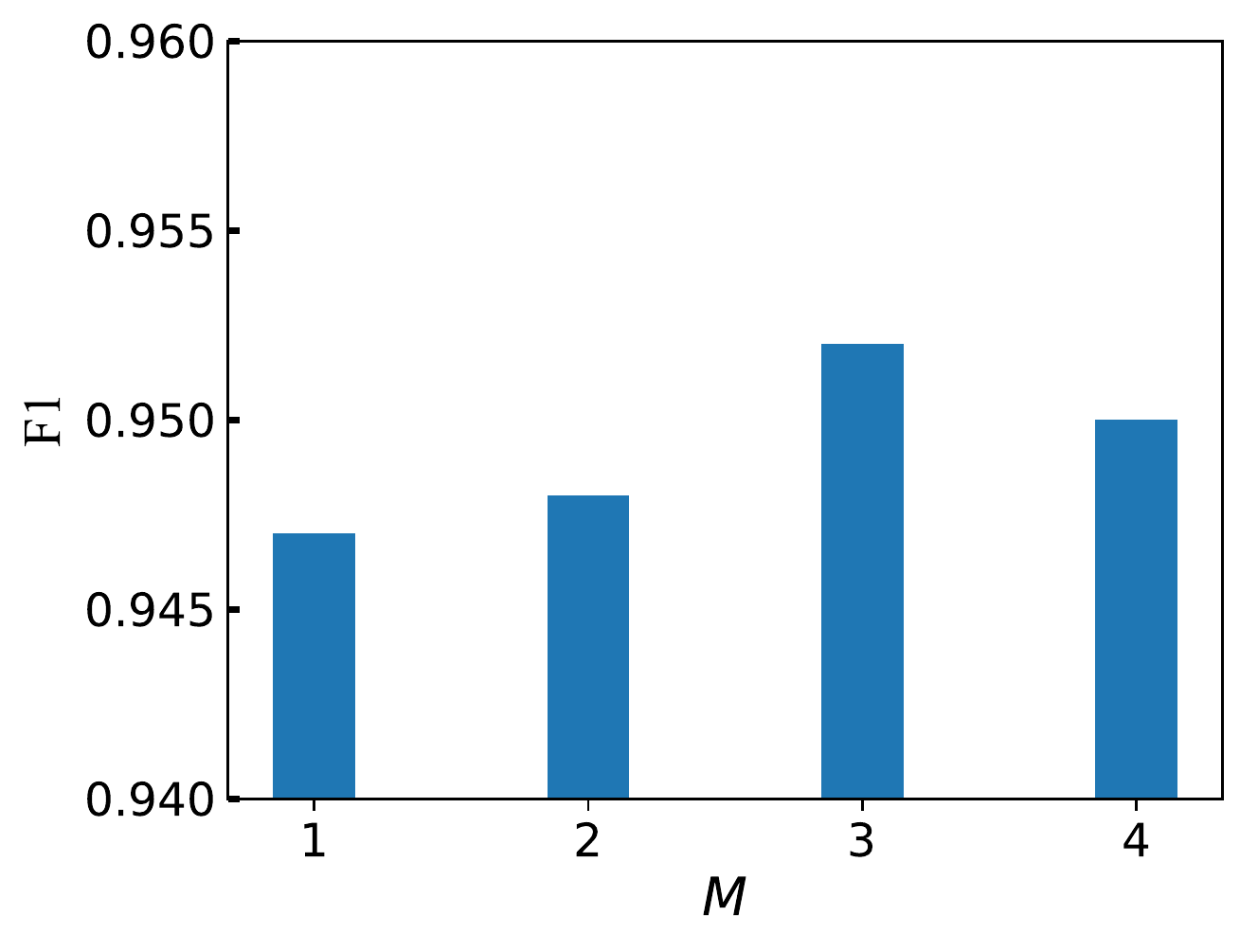}
		\label{fig:effect-facet-alpha-f1}
	}
	\subfigure[Bit.OTC - AUC]{
		\includegraphics[scale=0.45]{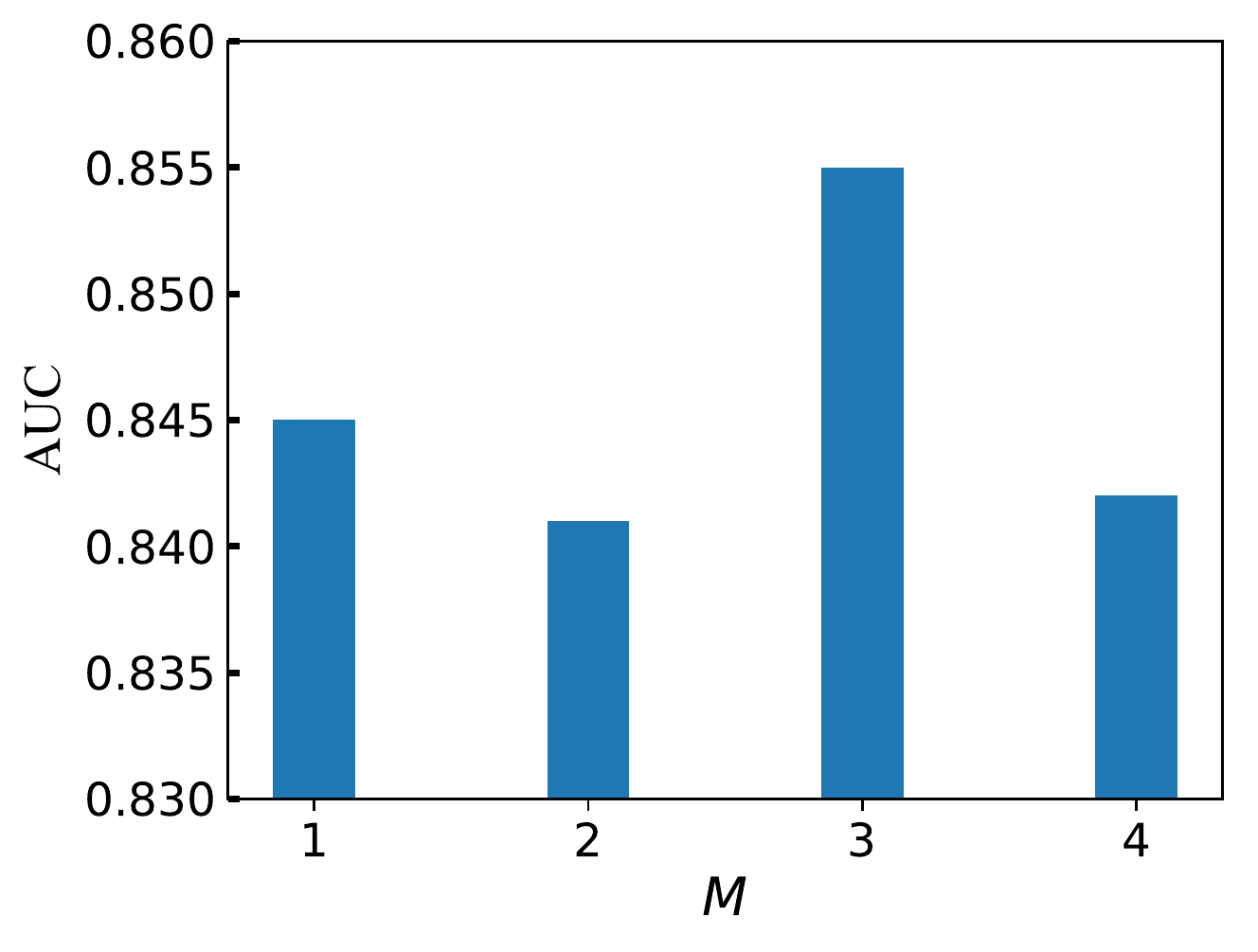}
		\label{fig:effect-facet-otc-auc}
	}
	\subfigure[Bit.OTC - F1]{
		\includegraphics[scale=0.45]{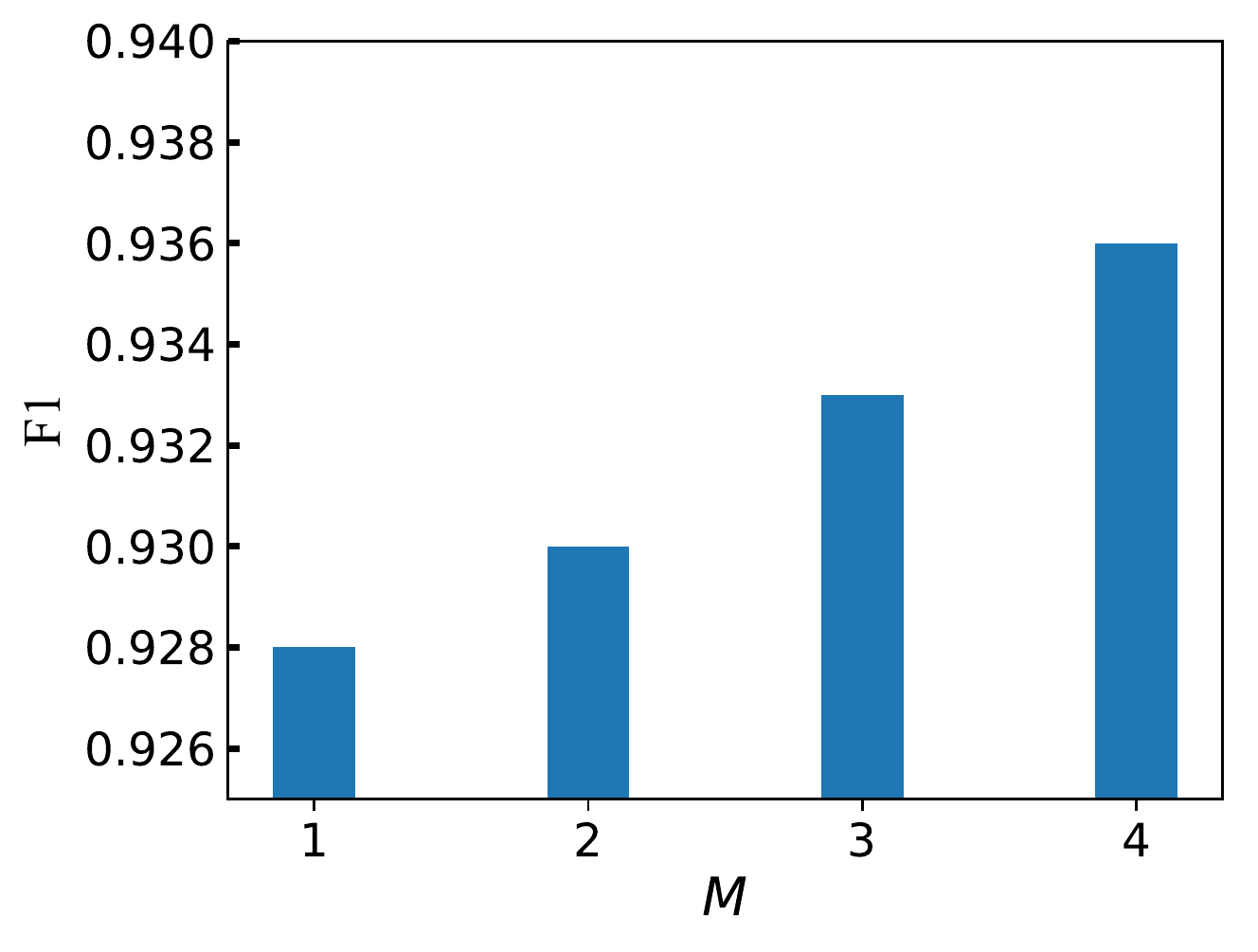}
		\label{fig:effect-facet-otc-f1}
	}
	\caption{The effect of different facet number $M$.}
	\label{fig:effect-facet-num-m}
\end{figure}

Furthermore, to explore the contributions of higher-order neighbors' aggregation, we also evaluate the following variant of our framework:

\begin{itemize}
    \item MUSE-1: It only aggregates information from the direct neighbors with multi-faceted attention.
\end{itemize}

\subsection{Experimental Settings}

In the experiment, we use the pre-divided training set and test set of the four datasets from ~\cite{li2020learning}. For each dataset, $80\%$ of the positive edges and $80\%$ of the negative edges are randomly selected as training set and the remaining is left as test set. The embedding dimension of each facet $D$ is fixed to 32, and the number of facets $M$ is set to 3. The number of orders $L$ is set to 2. The learning rate is set to $0.0001$. For the link sign prediction task, the embeddings of the source and target nodes are concatenated as the representation of links and then fed to a logistic regression classifier. All the  experiments are conducted using the same environment.

\subsection{Experimental Results}

\paragraph{Overall performance.} The overall performance of our propose MUSE framework and all baselines on the four real-world datasets is shown in Table \ref{tab:overall-performance}. The results demonstrate that MUSE, which utilizes muti-faceted attention and multi-order neighbor aggregation, outperforms all the baselines. The embeddings of nodes output by MUSE can obviously improve both the F1 and AUC metrics of the link sign prediction task. Specially, MUSE has a more significant improvement with regarding to AUC metric on the two social network sites (Slashdot and Epinions) than that on the two online bitcoin markets (Bitcoin-Alpha and Bitcoin-OTC). This is because users on the social network sites are more diverse and the formation of a positive or negative link are more complicated, in which the fine-grained mutli-faceted representation and attention is effective.

\paragraph{Effect of higher-order neighbor aggregation.}  Higher-order interactions between nodes are ubiquitous and play important roles in the direct link formation between nodes. For example, a seller and a buyer can conduct a transaction with the help of a broker or a broker platform, and the formation of a positive or negative link between the buyer and the seller maybe influenced by the higher-order interactions among them three. We evaluate the effect of higher-order neighbor aggregation by comparing the variant MUSE-1 with MUSE. From Table \ref{tab:overall-performance} we can see MUSE always performs better than MUSE-1, indicating the necessity of aggregation higher-order neighbors' information during signed network embedding.

\paragraph{Effect of facet number $M$.} The number of facets $M$ tunes the granularity of both the embeddings and the attentions. We explore the effect of facet number on the embeddings of nodes by comparing the performance of MUSE with different facet number $M$. For each $m \in [1, M]$, we obtain the embeddings of nodes using MUSE and the applied to the link sign prediction to calculate it F1 and AUC socres. The results is shown in Figure \ref{fig:effect-facet-num-m}. Taking an overall consideration of both F1 and AUC metrics, we can conclude too less facets cannot distinguish multiple characteristics of a node precisely while too many facets are redundant.

\begin{figure*}
	\centering
	\subfigtopskip=0pt
	\subfigure[Bit.Alpha - AUC]{
		\includegraphics[scale=0.45]{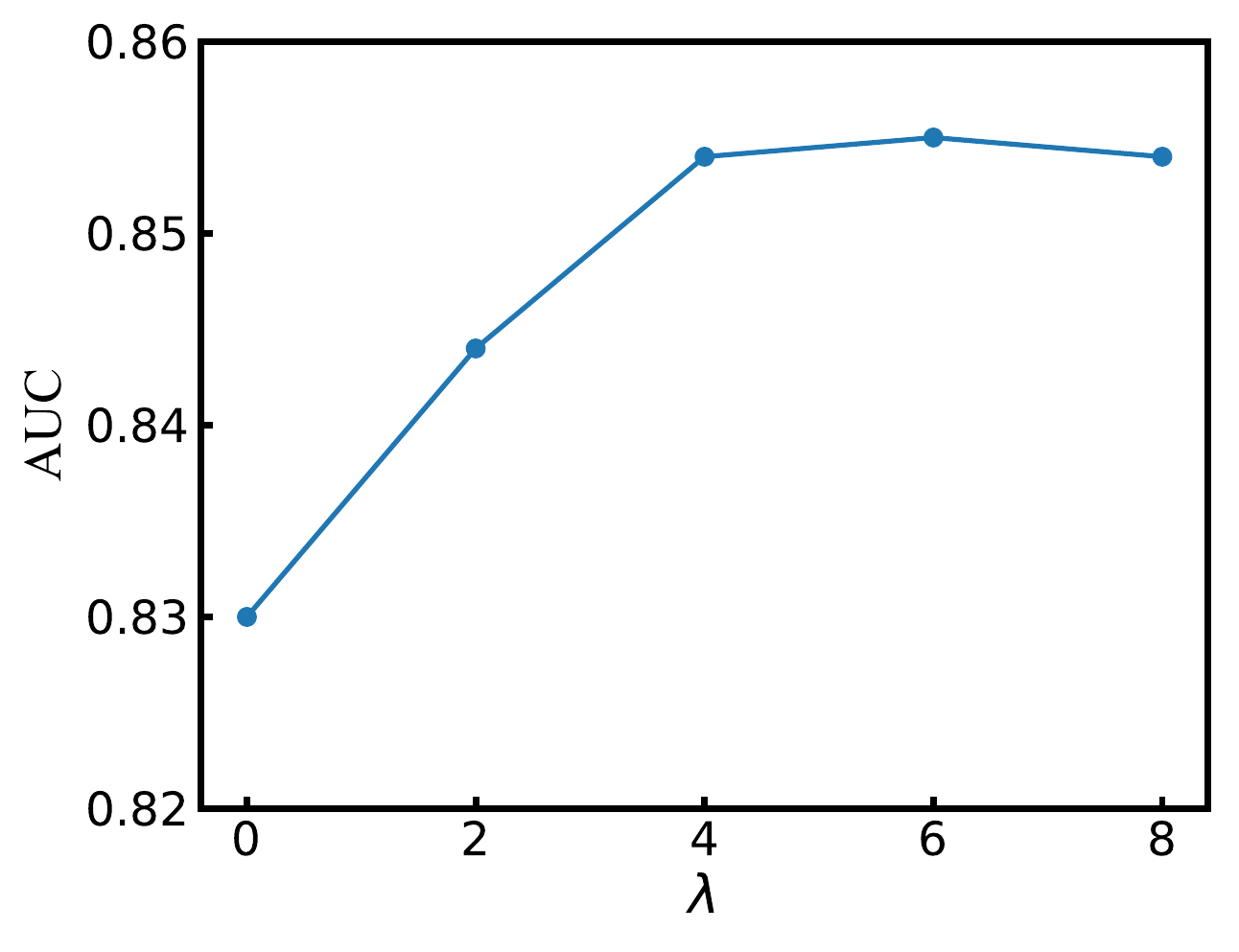}
		\label{fig:effect-loss-lambda-alpha-auc}
	}
	\subfigure[Bit.Alpha - F1]{
		\includegraphics[scale=0.45]{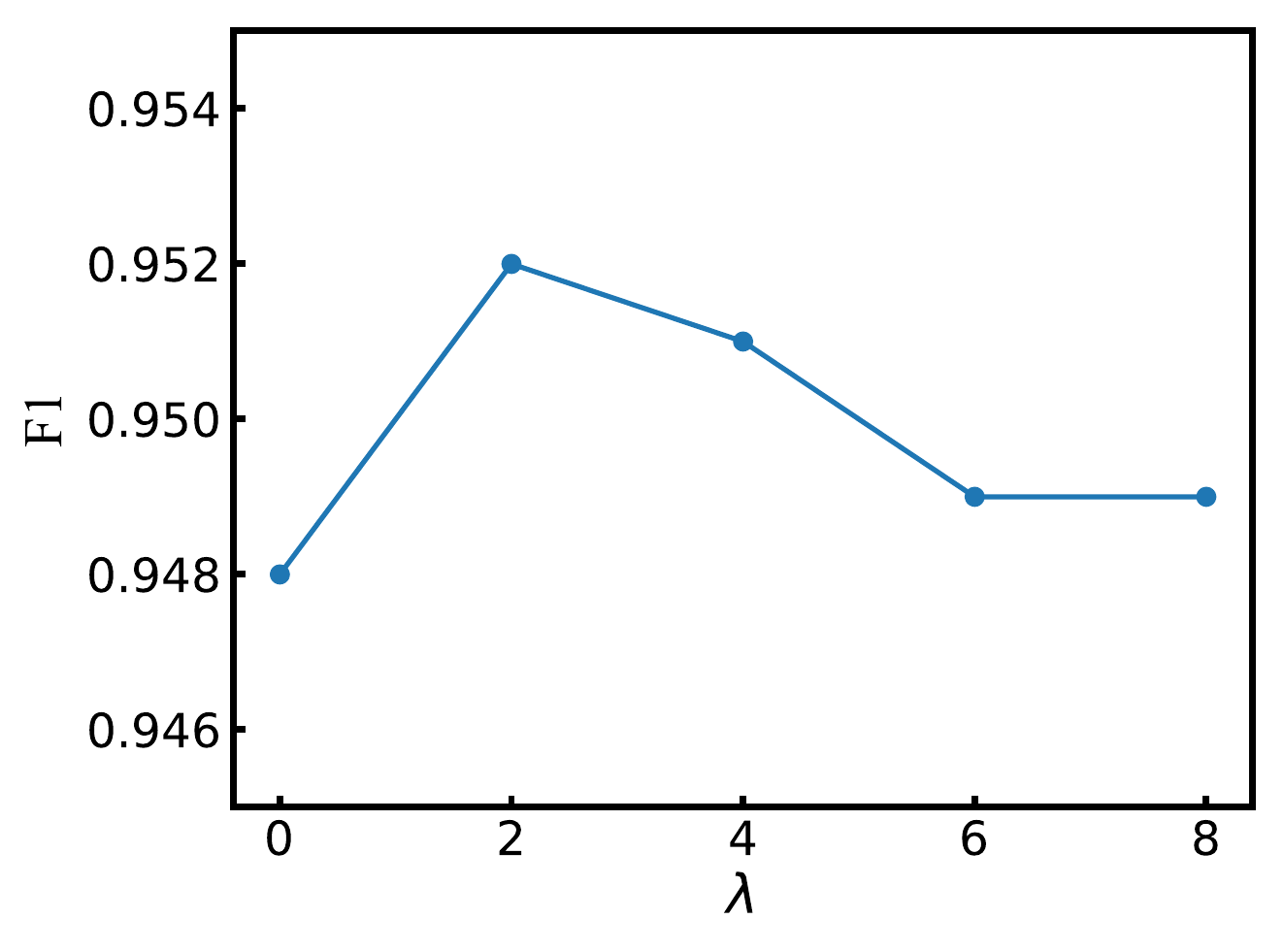}
		\label{fig:effect-loss-lambda-alpha-f1}
	}
	\subfigure[Bit.OTC - AUC]{
		\includegraphics[scale=0.45]{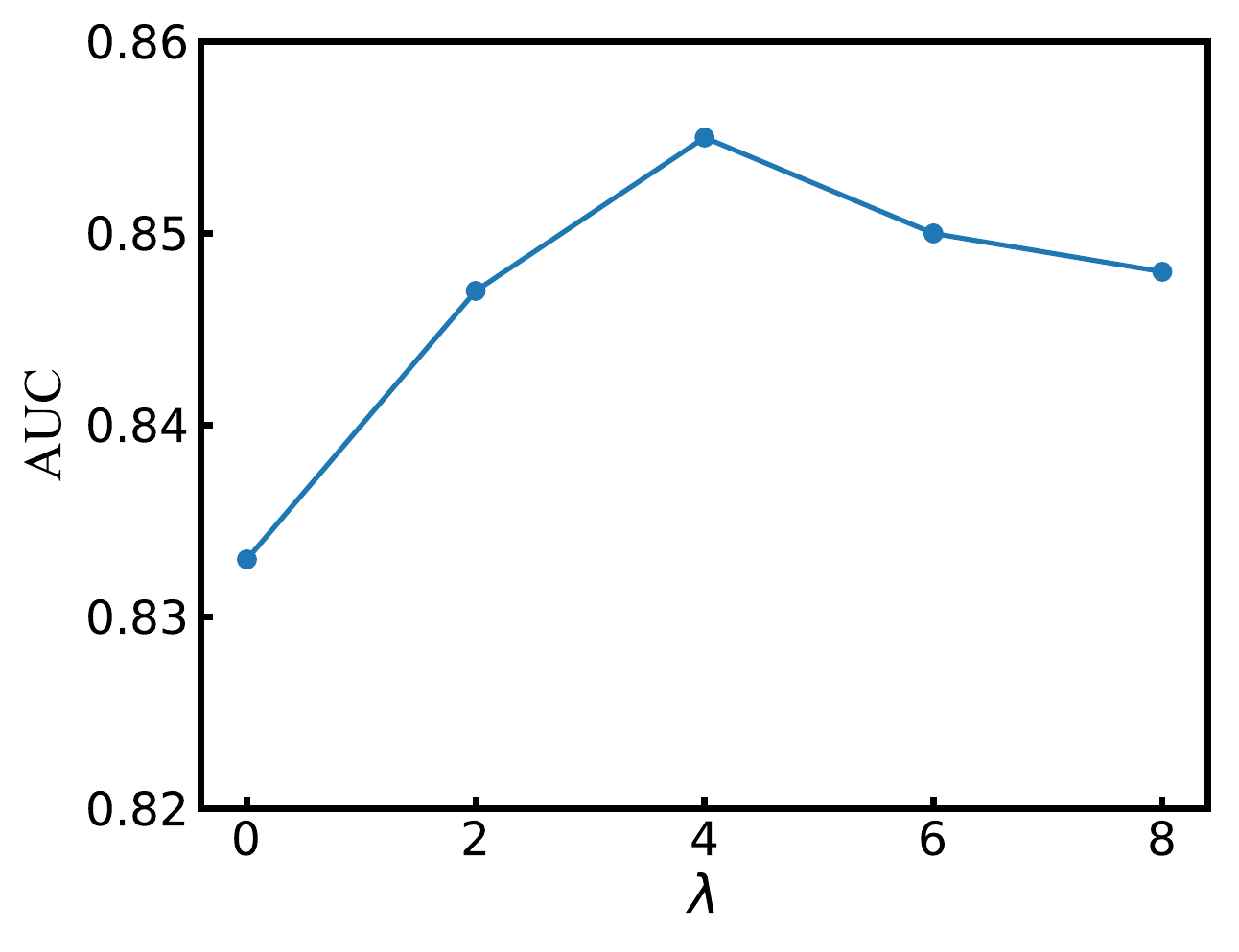}
		\label{fig:effect-loss-lambda-otc-auc}
	}
	\subfigure[Bit.OTC - F1]{
		\includegraphics[scale=0.45]{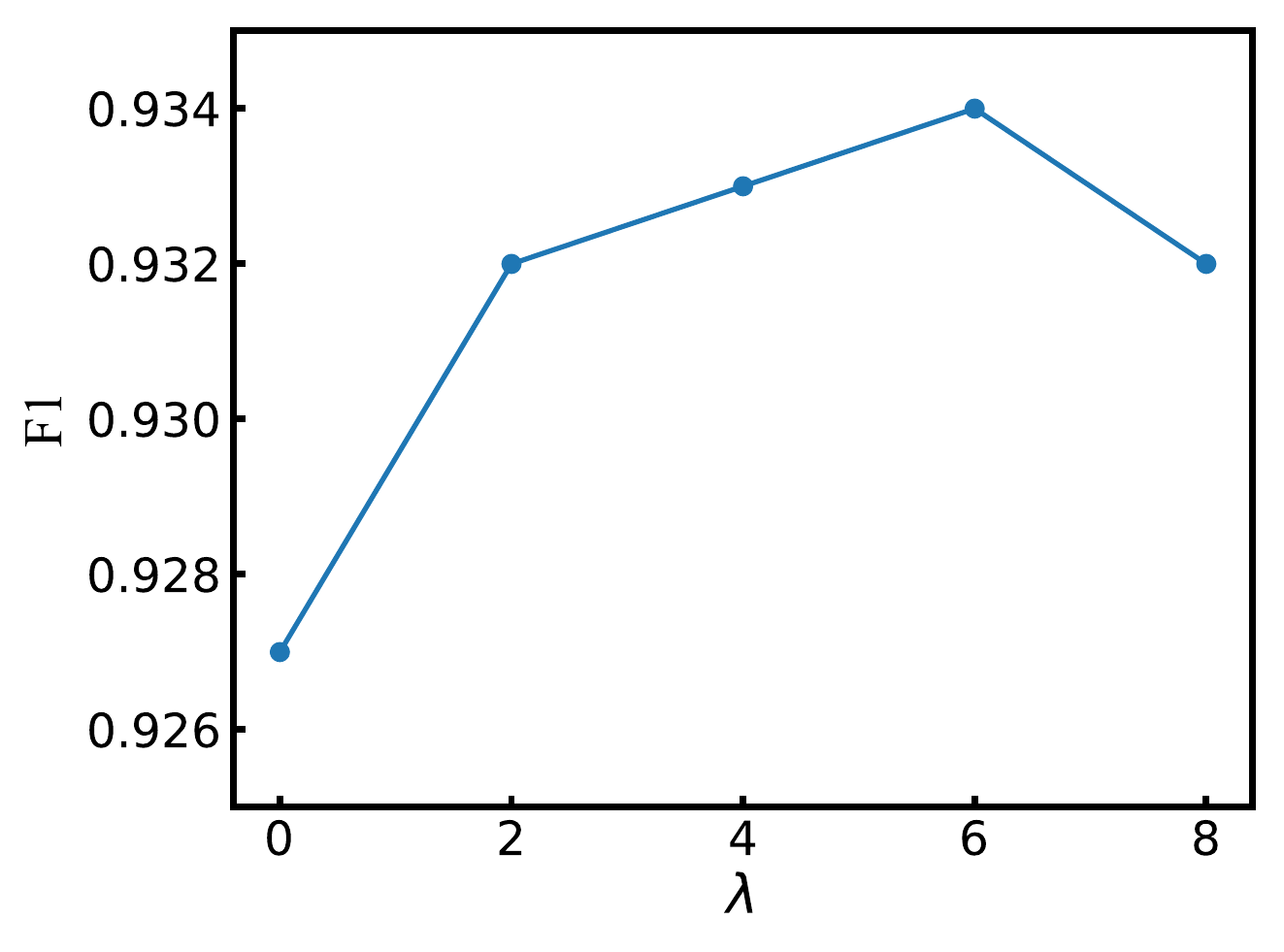}
		\label{fig:effect-loss-lambda-otc-f1}
	}
	\caption{The effect of loss balance $\lambda$.}
	\label{fig:effect-loss-lambda}
\end{figure*}

\paragraph{Effect of loss balance $\lambda$.} The loss balance $\lambda$ balances the importance between structure-preserving loss and link sign prediction loss. We evaluate the effect of loss balance $\lambda$ by increasing it from 1 to 8, and compare their different performance in datasets Bit.Alpha and Bit.OTC. The result is shown in Figure \ref{fig:effect-loss-lambda}, indicating either lower $\lambda$ or higher $\lambda$ will degrade the performance of MUSE. Lower $\lambda$ takes insufficient account for link type while higher $\lambda$ takes insufficient account for the network structure. There exists an optimal balance between these two optimization objectives, which is 4 in our experiment.

\section{Related Work}
In this section, we briefly discuss the most closely related work from two parts: signed network embedding and multi-faceted network embedding.

\paragraph{Signed network embedding.} Signed network analysis is attracting more and more attention from physics and computer science communities~\cite{tang2016survey}. Theories about signed networks is rooted in the domain of psychology and thereafter followed by the task of link sign prediction for application ~\cite{chiang2011exploiting} or signed centrality analysis ~\cite{kunegis2009slashdot} using traditional network analysis methods such as matrix factorization ~\cite{hsieh2012low,tang2016node}. With the prevalence of deep learning and network embedding, more and more research is focusing on signed network embedding, which aims to map nodes to a low-dimensional vector space to utilize machine learning models for signed network analysis. Signed network embedding methods usually use probabilistic models to preserve signed network structure or use graph neural network to aggregate information from neighbors. SNE ~\cite{yuan2017sne} utilizes a log-bilinear model to preserve signed network structure and optimizes the objective using maximum likelihood. Considering the directions of signed links, SIDE ~\cite{kim2018side} performs truncated random walks to obtain sampled pairs and adopts a likelihood optimization over both direct and indirect signed connections to obtain the node embeddings. Recently, various graph neural networks ~\cite{kipf2017gcn,velivckovic2017gat} are adopted to signed networks. SGCN ~\cite{derr2018signed} extends GCN ~\cite{kipf2017gcn} for unsigned network to signed networks and generates both balanced and unbalanced embeddings of a node simultaneously. To alleviate the problem that GCN cannot distinguish the different importance of neighbors during neighbor information aggregation, SiGAT ~\cite{huang2019signed} introduces attention mechanism and further SNEA ~\cite{li2020learning} improves it by utilizing a more effective masked self-attentional mechanism.

\paragraph{Multi-faceted network embedding.} The majority of existing network embedding methods only learn one single representation for each node, which is inefficient when modeling multiple facets of nodes. To overcome this issue, recent works ~\cite{epasto2019single,ParkYZKY020,LiuTLYZH19} have proposed to focus on obtaining multiple vector representations for each node in the network. For example, based on Deepwalk ~\cite{perozzi2014deepwalk}, PolyDeepwalk ~\cite{LiuTLYZH19} maximizes the likelihood of obtained observations by performing random walk. More precisely, PolyDeepwalk sets multi-faceted representations for nodes, and the facet of each node is sampled by the distribution known in advance. Furthermore, the unsupervised embedding method SPLITTER ~\cite{epasto2019single} divides nodes into multiple representations to make cluster structure easier to identify. MNE ~\cite{0002GC18} trains multiple embeddings based on matrix factorization, while the diversity is constrained by HSIC (Hilbert Schmidt Independence Criterion). However, multiple facets are not considered in existing signed network embedding methods.

\section{Conclusion and Future Work}

In this work, we propose a multi-faceted attention-based signed network embedding framework MUSE. The general idea behind the framework is that multiple facets do exist in the representation of nodes and the formation of links in signed networks, and adopting a multi-faceted attention will result in a finer granularity of nodes' embeddings. Specially, we design a multi-faceted attention mechanism to utilize both intra- and inter-facet attention for signed network embedding. Moreover, we also take into account the higher-order relations between node pairs and propose to aggregate information from multi-order balanced and unbalanced neighbors. Experiments on four real-world datasets demonstrate the effectiveness of our proposed MUSE framework As the structure of many real-world signed networks evolve over time and unbalanced triangles would evolve to balanced ones gradually, in the future, we plan to extend the multi-faceted attention (MFA) to dynamic signed networks.

\begin{acks}
This work is supported by the National Natural Science Foundation of China (Grant No. 61872002, U1936220), the University Natural Science Research Project of Anhui Province (Grant No. KJ2019A0037), the National Key R$\&$D Program of China (Grant No.2019YFB1704101) and the Anhui Provincial Natural Science Foundation (Grant No. 2008085QF307).
\end{acks}

\bibliographystyle{ACM-Reference-Format}
\bibliography{muse-ref}

\end{document}